\documentclass[sigconf]{acmart}

\usepackage{color}
\usepackage{algorithmic}
\usepackage{graphicx}
\usepackage{textcomp}
\usepackage{xcolor}
\usepackage{subfigure}
\usepackage{enumitem}
\usepackage{multirow}
\usepackage{pifont}
\usepackage[ruled,linesnumbered]{algorithm2e}
\usepackage{balance}

\newcommand\revise[1]{\textcolor{black}{#1}}

\AtBeginDocument{%
  }

\setcopyright{acmcopyright}
\copyrightyear{2023}
\acmYear{2023}
\acmDOI{XXXXXXX.XXXXXXX}

\copyrightyear{2023}
\acmYear{2023}
\setcopyright{acmlicensed}\acmConference[CF '23]{20th ACM International Conference on Computing Frontiers}{May 9--11, 2023}{Bologna, Italy}
\acmBooktitle{20th ACM International Conference on Computing Frontiers (CF '23), May 9--11, 2023, Bologna, Italy}
\acmPrice{15.00}
\acmDOI{10.1145/3587135.3592200}
\acmISBN{979-8-4007-0140-5/23/05}

\begin{document}

\title{DistSim: A performance model of large-scale hybrid distributed DNN training}

\author{Guandong Lu}
\orcid{0009-0000-6759-9108}
\affiliation{%
  \institution{Shanghai Jiao Tong University}
  \institution{Shanghai Qi Zhi Institusion}
  \city{Shanghai}
  \country{China}
}
\email{lugd0525@sjtu.edu.cn}

\author{Runzhe Chen}
\orcid{0009-0001-8955-2201}
\affiliation{%
  \institution{Shanghai Jiao Tong University}
  \institution{Shanghai Qi Zhi Institusion}
  \city{Shanghai}
  \country{China}
}
\email{runzhe_chen@sjtu.edu.cn}

\author{Yakai Wang}
\orcid{0009-0009-0835-6304}
\affiliation{%
  \institution{Shanghai Jiao Tong University}
  \institution{Shanghai Qi Zhi Institusion}
  \city{Shanghai}
  \country{China}
}
\email{wamgyakai@sjtu.edu.cn}

\author{Yangjie Zhou}
\orcid{0000-0002-3652-5437}
\affiliation{%
  \institution{Shanghai Jiao Tong University}
  \institution{Shanghai Qi Zhi Institusion}
  \city{Shanghai}
  \country{China}
}
\email{yj_zhou@sjtu.edu.cn}

\author{Rui Zhang}
\orcid{0009-0008-3887-9344}
\affiliation{%
  \institution{Huawei Technologies Co., Ltd}
  \city{Shenzhen}
  \country{China}
}
\email{zhangrui262@huawei.com}

\author{Zheng Hu}
\orcid{0000-0002-3526-0297}
\affiliation{%
  \institution{Huawei Technologies Co., Ltd}
  \city{Shenzhen}
  \country{China}
}
\email{hu.zheng@huawei.com}

\author{Yanming Miao}
\orcid{0009-0004-3244-2660}
\affiliation{%
  \institution{Huawei Technologies Co., Ltd}
  \city{Shenzhen}
  \country{China}
}
\email{miaoyanming@huawei.com}

\author{Zhifang Cai}
\orcid{0009-0000-4756-8806}
\affiliation{%
  \institution{Huawei Technologies Co., Ltd}
  \city{Shenzhen}
  \country{China}
}
\email{caizhifang@huawei.com}

\author{Li Li}
\orcid{0009-0005-6099-614X}
\affiliation{%
  \institution{Shanghai Jiao Tong University}
  \city{Shanghai}
  \country{China}
}
\email{lilijp@sjtu.edu.cn}

\author{Jingwen Leng}
\authornote{Jingwen Leng and Minyi Guo are the corresponding authors.}
\orcid{0000-0002-5660-5493}
\affiliation{%
  \institution{Shanghai Jiao Tong University}
  \institution{Shanghai Qi Zhi Institution}
  \city{Shanghai}
  \country{China}
}
\email{leng-jw@sjtu.edu.cn}

\author{Minyi Guo}
\orcid{0000-0003-0034-2302}
\authornotemark[1]
\affiliation{%
  \institution{Shanghai Jiao Tong University}
  \institution{Shanghai Qi Zhi Institusion}
  \city{Shanghai}
  \country{China}
}
\email{guo-my@cs.sjtu.edu.cn}

\renewcommand{\shortauthors}{G. Lu, R. Chen, Y. Wang, Y.Zhou, R. Zhang, Z. Hu, Y. Miao, Z. Cai, L. Li, J. Leng, and M. Guo}

\begin{abstract}
With the ever-increasing computational demand of DNN training workloads, distributed training has been widely adopted. A combination of data, model and pipeline parallelism strategy, called hybrid parallelism distributed training, is imported to tackle the problem of deploying large-scale models. However, how to evaluate the hybrid strategy and the utilization of each device remains a challenge since existing works either profile on a real large-scale cluster with high time and money costs or only analyze a specific type of parallelism without considering the hybrid parallelism. In this work, we proposed DistSim, an event-based performance model to accurately analyze each device's computation and communication activities with low profiling costs. DistDim breaks down the model into events according to the given distributed strategy, which can be profiled on two nodes. Then DistSim leverages the hierarchy of different parallel strategies to generate the computation and communication event-flow from layer level to model level and finally the activity timeline of each device participating in training. Experiment shows that DistSim can reach \revise{<4\%} errors when predicting distributing training batch time and \revise{<5\%} errors when predicting a single device's activity time in various hybrid strategy settings. We also provide a use-case of DistSim, automatically evaluate and search the best distributed training strategy, and find a hybrid strategy with at most $7.37\times$ throughput improvement. 
\end{abstract}

%




\maketitle

\section{Introduction}

Deep Neural Networks (DNNs) have been widely used in many areas and tasks due to their superior accuracy. 
For better accuracy, researchers have been proposing larger and deeper networks and the number of model parameters is increasing exponentially from 0.34 billion in Bert-Large ~\cite{bert} to 175 billion GPT-3~\cite{GPT-3}. It is impractical to deploy such large models on a single device. At the same time, since it is more difficult in convergence for a large-scale model, the time to train a model is increasing rapidly, from 4 days in Bert~\cite{bert} to 355 GPU years computed \revise{in} GPT-3~\cite{GPT-3}.

To address these problems, several solutions including designing domain specific accelerators~\cite{guo2020balancing, wang2021dual} and efficient computing operators~\cite{zhou2021characterizing, guo2022nesting,zhou2023ugrapher}, model pruning and sparsity~\cite{guo2020accelerating}, quantization~\cite{guo2022squant, guo2022ant}, and distributed training come into the picture. In distributed training, several distributed training strategies have been proposed for efficient training: data parallelism~\cite{DDP}, which increases the batch size to accelerate convergence; model parallelism~\cite{megatron} and pipeline parallelism~\cite{gpipe, dapple}, which partitions model into several pieces that can be deployed on a single device. Depending on the combination of these strategies, called hybrid parallelism distributed training~\cite{deepspeed}, large models are able to be trained in large-scale clusters with a reasonable speedup.

However, it still costs days to weeks to complete a training task~\cite{megatron-exp} even with the help of hybrid distributed training.
Not only hardware sharing~\cite{GPU-sharing} and resource management~\cite{resource-management, veltair} should be considered, it is also important to find the best training strategy in a large strategy search space. 
For a given distributed training strategy, it takes lots of time to initialize, partition model and profile several iterations to get the averaged throughput. When searching for a broad strategy space in large-scale clusters, it will cost a large amount of time and money before actual training to find such a solution.
It is difficult to design an accurate throughput or utilization estimation without profiling. Some prior works use analytical models to define the computation time~\cite{distIR, accpar}, i.e., operation count divided by computation capacity, but they have a large accuracy gap compared to the actual training process.
Meanwhile, research institutions or small companies usually do not own a large-scale cluster, which is required for evaluating their strategy-finding algorithms. As a result, they often need to rent clusters with additional expenses. For example, to rent a cluster with 2048 GPU devices, one needs to spend 7168 ($3.5 \times 2048$) USD per hour in an AWS cluster~\cite{aws}, before the actual training.


Therefore, it is important to have an accurate and cost-effective analysis tool to evaluate various distributed training strategies. Prior works like Daydream~\cite{Daydream} and dPRO~\cite{dPRO} try to address this challenge, but they only focus on computation and communication dependency in data parallelism. The complexity of hybrid parallelism distributed training has not been addressed yet.

In this work, we propose DistSim, an accurate simulator to analyze the performance of arbitrary distributed DNN training strategies in large-scale clusters. We dive into the relationships between each strategy in hybrid parallelism and develop our simulator with the following two insights.
First, there exists extensive profiling redundancy in hybrid parallelism. For example, each replica in data parallelism performs the same forward and backward computation; each device in model parallelism computes the same operator with the same input and weight shape in one layer. There is no need to profile all replicas as the actual training does.
Second, different parallelism strategy has different dependency level in hybrid distributed training, which we define as the hierarchical dependency of hybrid parallelism. Each strategy has its own data dependencies and is unrelated to others. For example, changing the partition size of model parallelism does not affect on the dependencies in pipeline parallelism since pipeline parallelism is unrelated to the computation inside one layer -- where model parallelism participates in.

Based on these insights, we define ``event'' to eliminate the redundancy of profiling computation and communication in hybrid parallelism. Without loss of generality, in clusters with homogeneous devices and no network hierarchy, the same computation and communication performed by different devices can be gathered into one event and need to be profiled only once. Hence the profiling of the whole training process in large-scale clusters can be reduced to a minimal number of 2 nodes (1 node is unable to profile the inter-node communication). 
All the events are used to construct a complete training timeline from the lowest level -- model parallelism, to the highest level -- data parallelism.

DistSim models the training process with the event-based construction. We evaluate the simulation accuracy of DistSim and show that DistSim is able to estimate the performance of different hybrid training strategies accurately. In addition, we provide a simple use case of DistSim: finding a better hybrid training strategy.

In summary, we make the following contributions:

\begin{itemize}[leftmargin=*]
\vspace*{-0.1cm}
    \item We identify the profiling redundancy in hybrid distributed training and define a concept of the event, which avoids redundant computation and communication, and reduces the profiling overhead from real large-scale clusters to only two nodes.
    \item We build DistDim, a performance model leveraging the hierarchy of different distributed training strategies to construct the timeline of each device using profiled events. We propose detailed modeling methods for data, pipeline, and model parallelism. DistSim has the ability to deal with any combination of these strategies in hybrid parallelism.
    \item We evaluate DistSim from different aspects and show that DistSim can evaluate different hybrid training strategies accurately both in batch-time -- critical path (\revise{<4\%} error) and in single device's activity (\revise{<5\%} error). We also provide a use case of DistSim, searching for the best parallel strategy and finding a hybrid strategy with at most $7.37\times$ throughput improvement in an unseen 48-layer model ``Bert-exLarge''.
\vspace*{-0.1cm}
\end{itemize}
\section{Background and Motivation}

This work presents an accurate and cost-effective performance model to analyze training time and device activities in distributed DNN training.
It allows users to keep track of the detailed information of every device given a model to train and hybrid training strategies to apply. 
In this section, we first present a brief background of different distributed training strategies.
We then present the motivation that such a new performance model is necessary.
Specifically, we explain why existing performance models, including direct profiling or analytical model, are ineffective and insufficient.


\subsection{Hybrid Distributed DNN Training}

\begin{figure}[t]
\vspace*{-0.2cm}
\centerline{
  \subfigure[data parallelism]{
     \label{dp}
     \includegraphics[width=\linewidth]{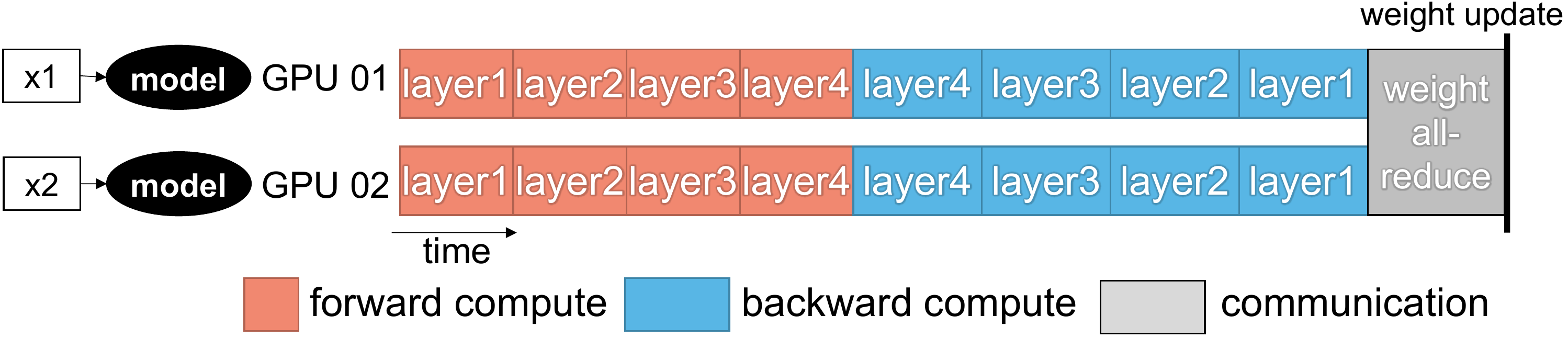}
  }
}
\vspace*{-0.3cm}
\centerline{
  \subfigure[model parallelism]{
     \label{mp}
     \includegraphics[width=\linewidth]{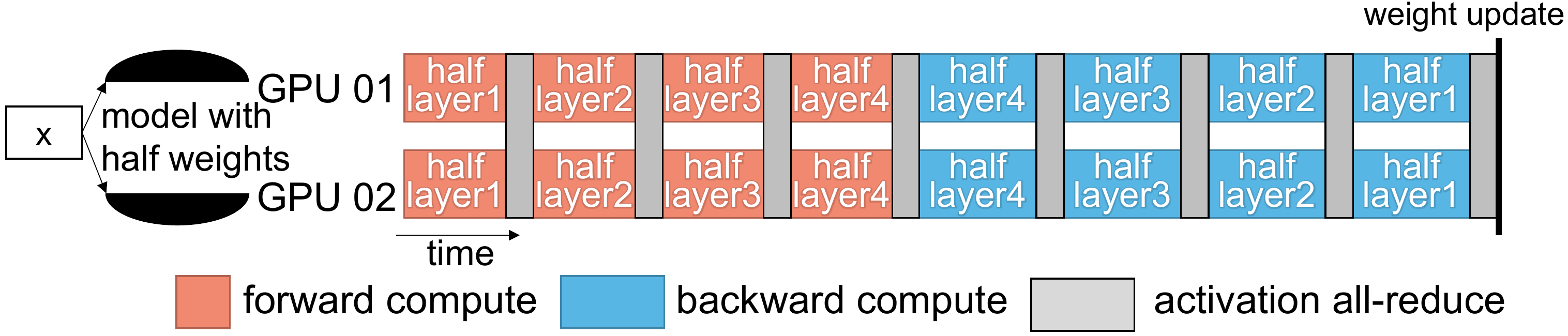}
  }
}
\vspace*{-0.3cm}
\caption{\revise{Examples of data and model parallelism.}}
\vspace*{-0.4cm}
\label{data_model}
\end{figure}

To train a large-scale DNN models, prior works have exploited various  parallelism as follows.

\subsubsection{Data Parallelism.\label{bg_dp}}
In data parallelism, every device holds the same model. Input data is partitioned among batches and distributed into devices. Each device independently computes the forward and backward steps, and the gradient is gathered to sync the model's state. Fig.~\ref{dp} shows a brief timeline of data parallelism. Using data parallelism for DNN training increases the batch-size of the training process, which also speeds up the training with a large learning rate. As such, data parallelism is widely used.

There are two variants of data parallelism: Parameter-Server (PS)~\cite{parameter-server} and Ring-AllReduce~\cite{ring-allreduce}. In PS method, workers send their gradients to parameter servers and receive the latest state of the model. In Ring-AllReduce, all devices send a part of the gradients to aggregate, then broadcast its aggregation result. 
In this work, we mainly discuss and model ring-AllReduce-based data parallelism, as they are widely used by distributed training frameworks such as PyTorch Distributed~\cite{DDP} and Horovod~\cite{horovod}.

\subsubsection{Model Parallelism.}
\revise{Model parallelism is also named as tensor model parallelism by Megatron-LM~\cite{megatron-exp} or horizontal parallelism by DistIR~\cite{distIR}.} First used by training AlexNet~\cite{alexnet}, model parallelism splits the weight matrix into multiple devices. Each device computes with its partitioned weight matrix for the same input and gathers the partial output with other devices, as shown in Fig.~\ref{mp}.
Model parallelism is useful when device memory is insufficient to deploy a large-scale model. Recently, Megatron-LM~\cite{megatron-exp} has implemented model parallelism into distributed training frameworks for partitioning large language processing models.

\subsubsection{Pipeline Parallelism.}
\revise{Although pipeline parallelism is sometimes also called inter-layer model parallelism or pipeline model parallelism~\cite{megatron}, pipeline is the main concept to distinguish from the previously introduced model parallelism.} Pipeline parallelism partitions models into layer-wise stages and deploys them into different devices. Devices only need to compute the stages in charge and send their output activation to the next stage. 

The naive pipeline parallelism, only computing a layer once per iteration, reaches a low device utilization since worker has to wait until its input is prepared. This phenomenon is called pipeline bubbles. Prior work introduces micro-batch to overlap training and proposes various algorithms to reduce bubbles, which can be categorized into synchronous and asynchronous pipeline parallelism.

\begin{figure}[t]
\vspace*{-0.3cm}
\centerline{
  \subfigure[GPipe]{
     \label{gpipe}
     \includegraphics[width=\linewidth]{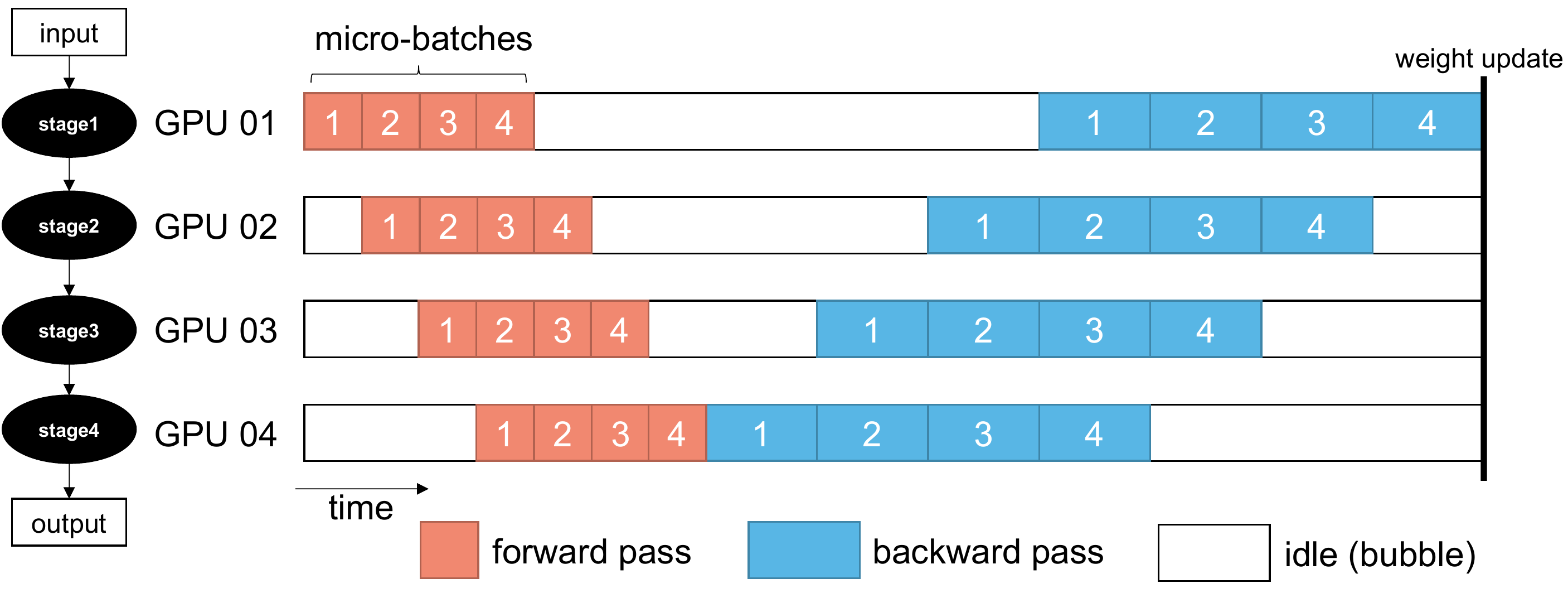}
  }
}
\vspace*{-0.3cm}
\centerline{
  \subfigure[Dapple]{
     \label{dapple}
     \includegraphics[width=\linewidth]{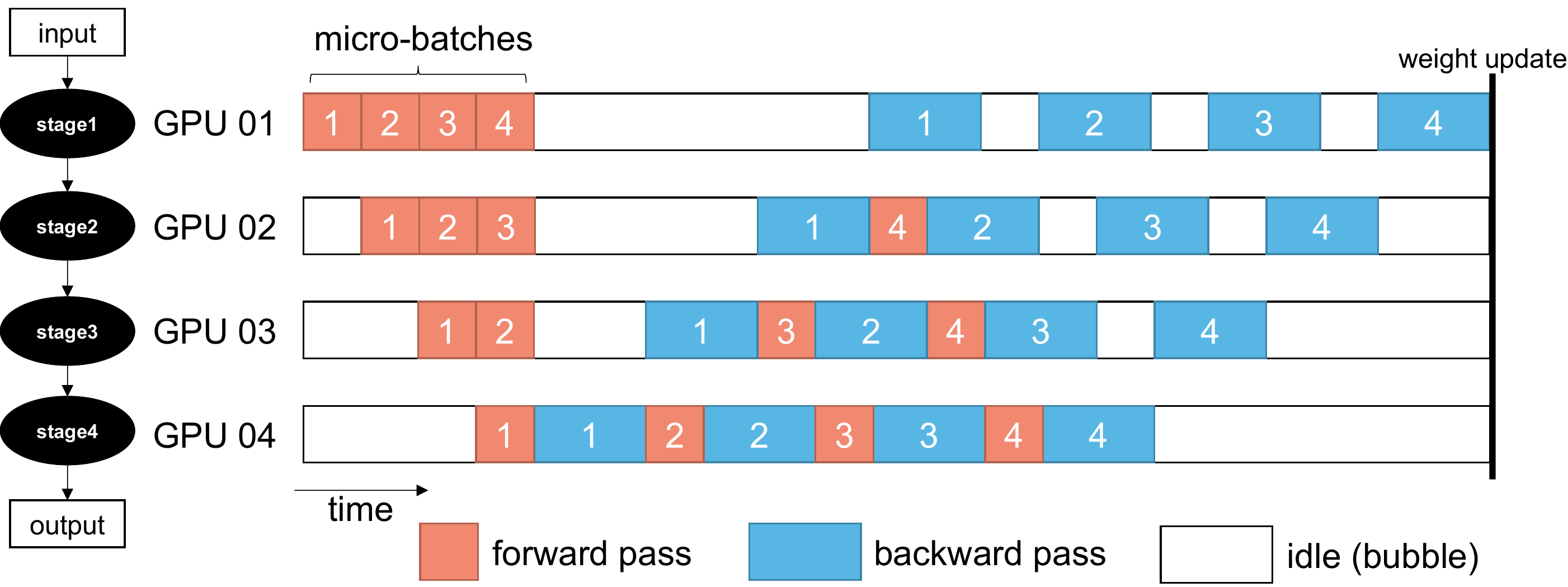}
  }
}
\vspace*{-0.3cm}
\caption{An example of two algorithms in pipeline parallelism to reduce pipeline bubbles: GPipe (a) and Dapple (b). The number in each stage represents micro-batch index. }
\vspace*{-0.4cm}
\label{pipeline}
\end{figure}

Synchronous pipeline parallelism has a synchronized weight updating phase for all devices to guarantee convergence. Unlike the naive pipeline, the training batch is divided into  micro-batches ~\cite{gpipe}, which can be computed in a pipelined approach. Prior works, such as GPipe ~\cite{gpipe}, Dapple ~\cite{dapple} and Chimera ~\cite{chimera} introduce various scheduling algorithms to increase utilization and reduce memory consumption. Fig.~\ref{pipeline} shows a brief example of GPipe (a) and Dapple (b), which explains how these algorithms overlap computing for micro-batches to reduce bubbles.

Asynchronous pipeline parallelism removes the universal weight-updating phase along devices. Each device updates its own weight and computes different micro-batches simultaneously, which reaches a high device utilization. However, although previous experiments~\cite{pipedream} show harmlessness, the lack of universal synchronization to update weight may affect training convergence. Thus we focus only on discussing synchronous pipeline parallelism in this work.

\subsubsection{Hybrid Parallelism.}
Prior work OWT~\cite{OWT} is the first to use hybrid parallelism, which is not strictly a new category of distributed training strategies, but partitions the model into more than one aspect and leverages the advantages of different strategies of data, model, and pipeline parallelism.
The representative work of hybrid parallelism includes PTD-P from Megatron-LM~\cite{megatron} and 3D-parallelism in DeepSpeed~\cite{deepspeed}.

\subsection{Why Not Profiling Performance Directly?}

To model performance, the most straightforward solution is to actually run and profile the training process on a real cluster to get the iteration time or devices' activities. However, this solution has a shortcoming: users must have access to the \textit{real} training cluster, which could be time- and money- consuming. For example, when researchers want to profile and analyze the utilization of devices in a particular hybrid training strategy with 2048 GPUs, which is a typical setting in Megatron-LM~\cite{megatron-exp}, they have to find such a 2048-device cluster to profile. Such cluster is uncommon to be seen and is very expensive to rent ($3.5 \times 2048 = 7168$ USD per hour in an AWS cluster~\cite{aws}). Unlike profiling on a single device, the high overhead in clusters makes it hard to directly profile, which is the first challenge of performance modeling.

\subsection{Why Not Analytical Model?}

Another way for performance modeling is to evaluate analytically with the information of training hardware and DNN training workload. For example, one common heuristic approach as prior work does~\cite{distIR}\cite{accpar} is: using the division of floating-point operators count and hardware computing capacity (FLOPS) to represent computation time and regarding the division of data transmission size and the bandwidth as the communication time. One problem of the heuristic approach is accuracy, as it is not often true that the DNN operators will full-utilize the hardware's computing resource~\cite{utilization}. 

\begin{figure}[t]
\centerline{\includegraphics[width=0.8\linewidth]{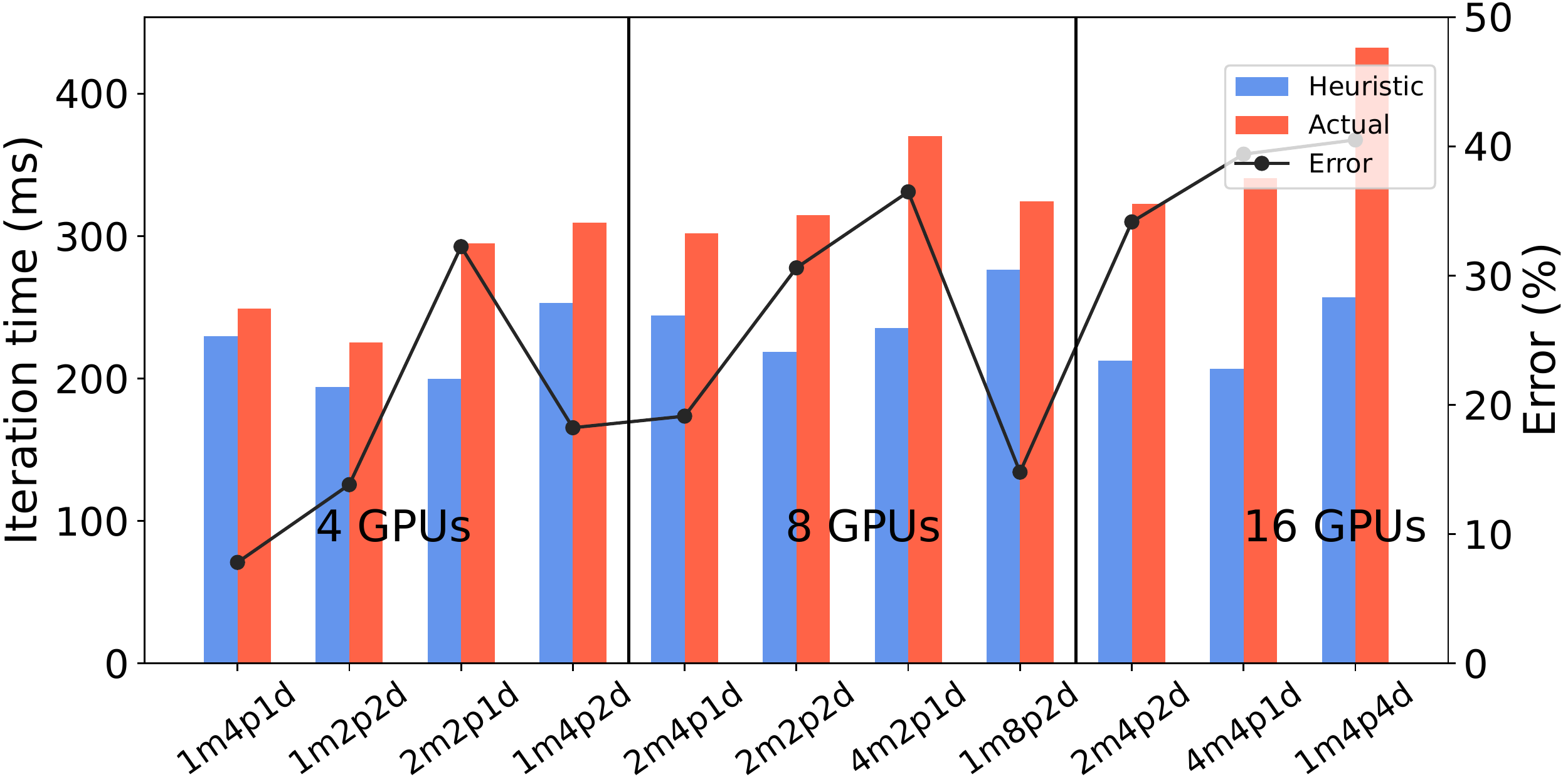}}
\vspace*{-0.1cm}
\caption{\revise{The iteration time comparison between analytical model and actual profiling result on the training of Bert-Large\cite{bert} with 4 to 16 A40 GPUs. The x-axis shows the notations of different distributed training strategies, which is defined in Sec.\ref{exp-setup}.}}
\vspace*{-0.3cm}
\label{res_heuristic}
\end{figure}

To show this problem, we evaluate the heuristic approach with Bert-Large~\cite{bert} model and 4-16 A40 GPUs with different distributed training strategies. The comparison of iteration time (from which the throughput of the training process can be conducted) presented in Fig.~\ref{res_heuristic} shows a significant bias from heuristic to real profiling, with at most 40.4\% error and 26.1\% on average. The high divergence of the heuristic evaluation makes results unreliable for further analysis, such as analyzing per-GPU utilization.

\subsection{Why Not Using Current Distributed Training Simulator?}

\begin{table}[b]
\Huge
\centering
\vspace*{-0.3cm}
\caption{\revise{A summary and comparison of different ways to model performance of large-scale DNN training, to show our motivation to a general and light-weighted modeling approach}}
\vspace*{-0.3cm}
\label{tab:motivation_comp}
\resizebox{\columnwidth}{!}{%
\begin{tabular}{|c|c|c|c|c|c|}
\hline
Methodology                       & Representative work & \begin{tabular}[c]{@{}c@{}}Support \\data \\ parallelism\end{tabular} & \begin{tabular}[c]{@{}c@{}}Support \\ model/pipeline\\ parallelism\end{tabular} & \begin{tabular}[c]{@{}c@{}}Low\\ profiling\\ overhead\end{tabular} & \begin{tabular}[c]{@{}c@{}} Accu-\\racy \end{tabular} \\ \hline
Direct profiling                  & Megatron-LM~\cite{megatron-exp}         & o                                                                   & o                                                                             & ×                                                                & o        \\ \hline
Analytical Model               & DistIR~\cite{distIR}, accPar~\cite{accpar}      & o                                                                   & o                                                                             & o                                                                & ×        \\ \hline
\multirow{2}{*}{Simulation-based} & daydream~\cite{Daydream}, dPro~\cite{dPRO}      & o                                                                   & ×                                                                             & o                                                                & o        \\ \cline{2-6} 
                                  & DistSim (Ours)       & o                                                                   & o                                                                             & o                                                                & o        \\ \hline
\end{tabular}%
}
\end{table}

A third possible approach is using current distributed training simulators such as Daydream~\cite{Daydream} and Dpro~\cite{dPRO}, as they can profile kernels and construct the training process without running them. Both simulators are constructed based on a critical assumption: tasks in distributed DNN training workloads are highly sequential~\cite{Daydream}, which means when one device finishes computing the first layer, it will naturally launch the second layer. This assumption enables them to ignore the dependencies among layers. Although the assumption is true in data parallelism, it is not always correct for other parallelism strategies. For example, after one device finishes computing the first layer, it may switch to the next micro-batch of pipeline parallelism and compute the first layer again, which breaks the computing sequence in DNN training. Therefore, these simulators can replay the training process of data parallelism, but cannot support other strategies in hybrid parallelism. We regard this problem as the complexity of dependencies in hybrid parallelism, which is shown as the second challenge of modeling.

\paragraph{Summary.} Prior approaches to model performance of large-scale hybrid distributed DNN training have their shortcomings. Particularly, because of the unsatisfying accuracy, it is not suitable to use an analytical approach to model performance, and profiling is the best choice. Directly profiling and using current profiling-based simulators will face the challenges of profiling consumption and the complexity of dependencies. Therefore there is a need for a \revise{light-weight} and generalized approach: profile on small-scale clusters and extrapolate towards the full-scale training process, which is what we proposed in this work, presented in Table~\ref{tab:motivation_comp}.

\section{DistSim}

Our method to model large-scale distributed training performance is based on two observations. In this section, we first introduce these observations, and then show how DistSim solves the challenges of profiling and dependency raised in the previous section.

\subsection{Observations}

\begin{figure}[tbp]
\centerline{\includegraphics[width=\linewidth]{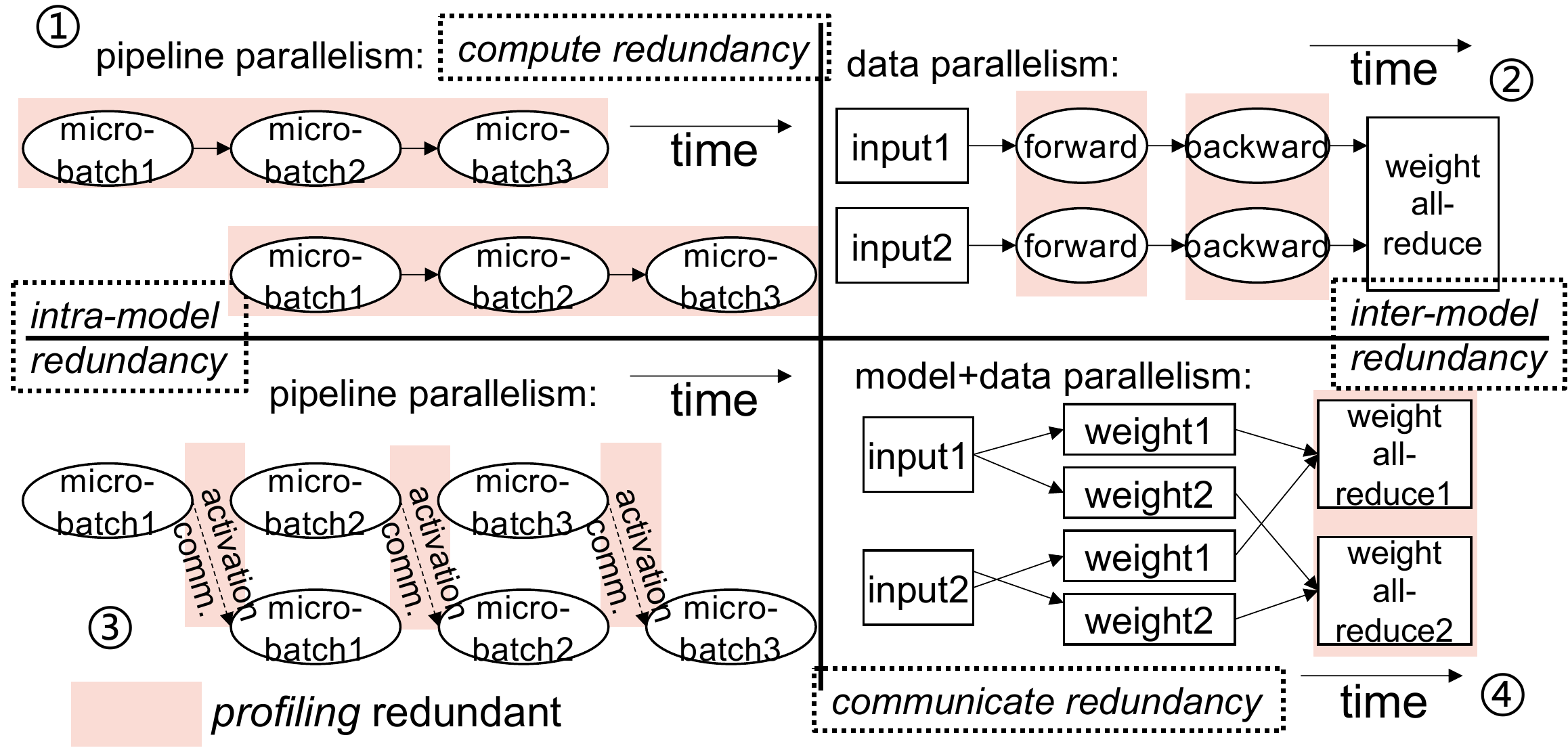}}
\vspace*{-0.3cm}
\caption{\revise{An illustration for profiling redundancy in hybrid training, including: same computation among micro-batches (\ding{172}) and different devices (\ding{173}), as well as transferring activations whose shape is the same (\ding{174}) and weights with same size (\ding{175}).}} 
\vspace*{-0.3cm}
\label{profiling_redundancy}
\end{figure}

\noindent\textbf{Observation 1: Profiling redundancy.} While training with data parallelism, certain devices perform the same computation since they all correspond to the same model.
One device may compute the same layer for different micro-batches in pipeline parallelism. The computation and communication redundancy are shown in Fig.~\ref{profiling_redundancy}. In addition, some devices may be idle, waiting for input from other devices (recall pipeline bubbles in Sec.~\ref{bg_dp}). Those repeated computing time or idle time will be profiled multiple times in direct training, which is unnecessary.

\begin{figure}[tbp]
\centerline{\includegraphics[width=\linewidth]{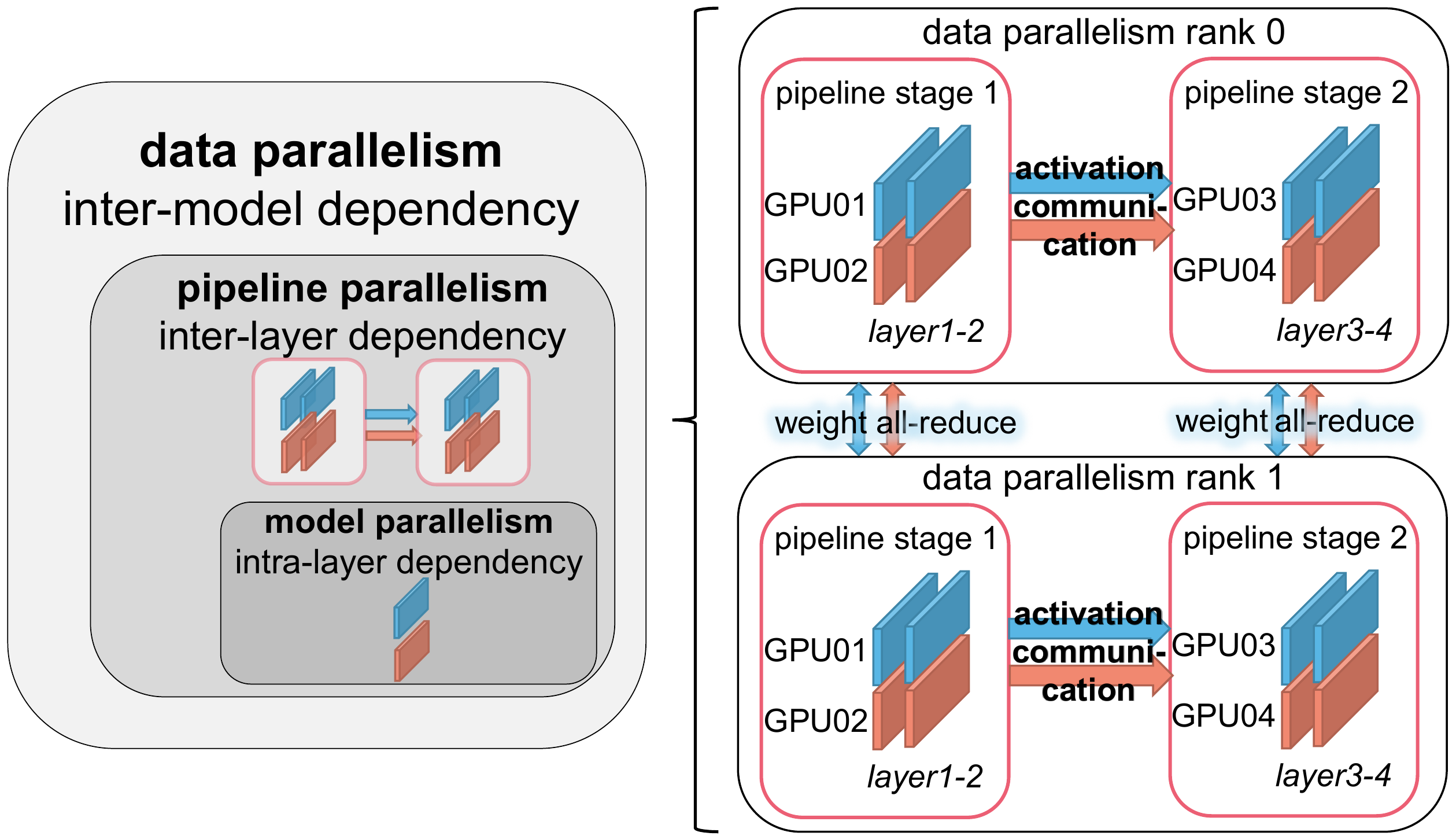}}
\vspace*{-0.2cm}
\caption{\revise{An illustration for hierarchical dependencies in hybrid distributed training. Data parallelism focuses on the weight dependencies across models among data parallelism. The dependencies within a model is pipeline parallelism in charge, and finally, they are controlled by model parallelism in one layer.}} 
\vspace*{-0.3cm}
\label{hierarchical_dependency}
\end{figure}

\noindent\textbf{Observation 2: Hierarchical dependency of different parallel strategies.} Different parallel strategies need to synchronize or communicate for different data, which is shown in Fig.~\ref{hierarchical_dependency}: data parallelism synchronizes weights, model parallelism gathers partial output, and pipeline parallelism communicates activation for different stages. Based on this fact, every strategy only focuses on its own dependency and ignores others'. This observation reduces the problem of handling dependence complexity to finding a way to combine each strategy's dependency together.

\subsection{Overview of DistSim}

\begin{figure}[tbp]
\centerline{\includegraphics[width=\linewidth]{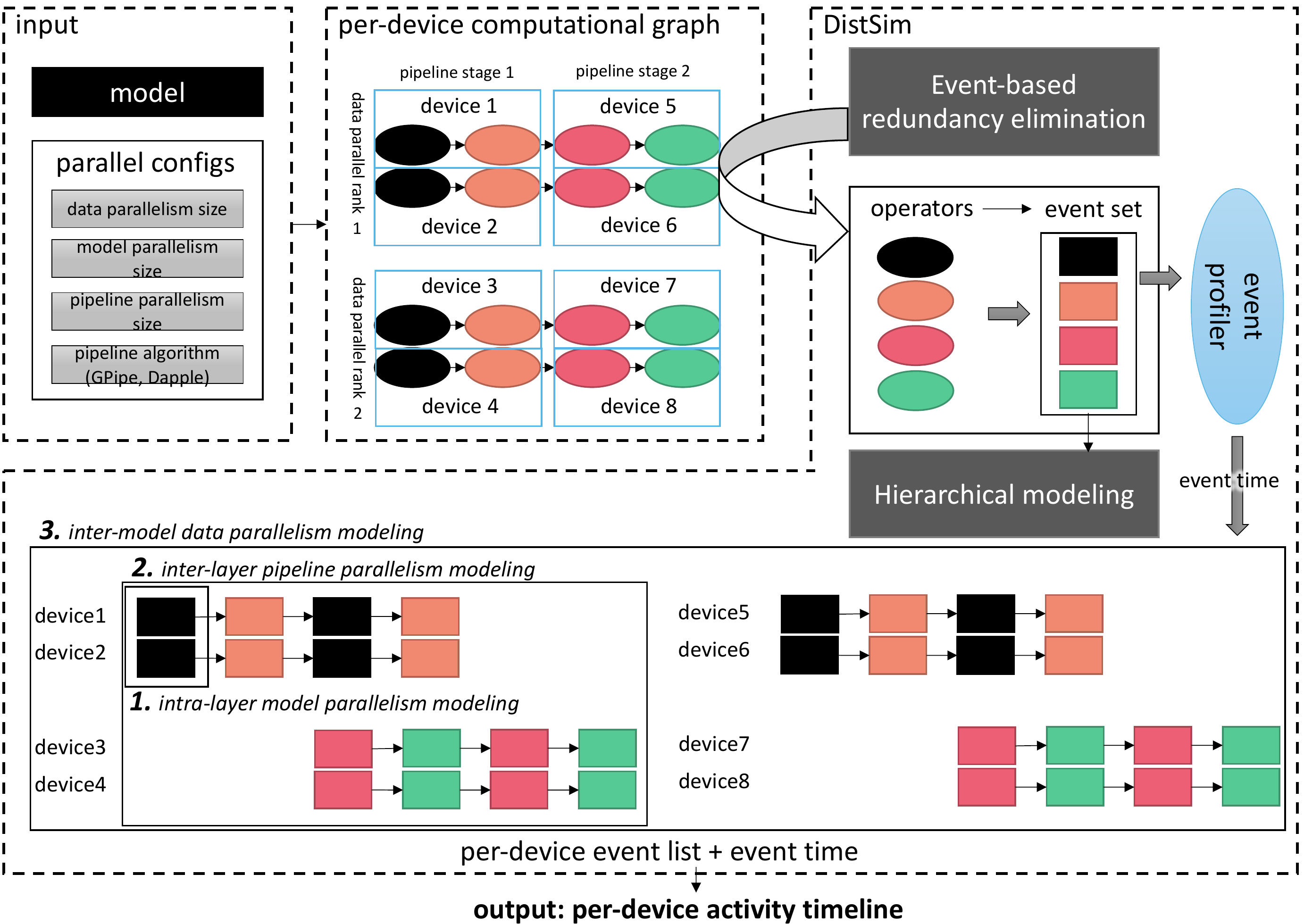}}
\vspace*{-0.2cm}
\caption{An overview of DistSim, which takes the model and distributed training configurations as input. DistSim first generate an abstraction called events from sub-models, then profiles and uses events to construct and recover the whole training process.} 
\vspace*{-0.2cm}
\label{distsim_overview}
\end{figure}

Fig.~\ref{distsim_overview} is an illustration of how DistSim works. Given a model and a particular configuration of parallel strategy, DistSim reduces the redundancy of profiling (\textit{Observation 1}) by defining an abstraction named \textit{event} to discover and preserve identical operators. As these events are essentially operators deployed on individual devices, they can be profiled only once and without large-scale clusters.
These events, which can be divided into computation and communication events based on the type of operators, are profiled by DistSim separately according to events' type.

After profiling and getting the detailed time information of events, DistSim recovers the timeline of the original training process by further defining and combining \textit{composed-event}, which contains several events. Composed-event aims to make each distributed strategy only focus on its own dependency (\textit{Observation 2}) when modeling. Each strategy's modeling constructs an event list differently (saved in composed-event) and DistSim combines these event lists together. The final timeline can be generated when combining these event lists with profiled event times.

The output of DistSim is a detailed execution timeline for the full-scale distribution training, which contains when and which device will compute and communicate for certain operators. The timeline can be used by users for further analysis, such as iteration-time, device utilization, and pipeline bubble analysis.

\revise{In DistSim, we profile to get the correct events' elapsed time. When users do not have a profiling device or want a light-weight and profiling-free way by sacrificing some accuracy, they can alternately use GPU simulators such as MGPUSim~\cite{MGPUsim} and operator predictors such as Habitat~\cite{Habitat}. In addition, the events' time can be stored and reused when modeling a new parallelism strategy as long as the model can generate the same event.}

\section{Framework Design Detail}

In this section, we describe the details in DistSim that complete the process in Fig.~\ref{distsim_overview}, including (i) generating events; (ii) profiling events; (iii) hierarchical modeling based on these events.

\subsection{Event Generator}

We leverage the model partition function in current distributed training frameworks such as PyTorch Distributed~\cite{DDP}. During actual training, with the model and parallel strategy configurations inputted, frameworks will generate per-device sub-models to deploy on each distributed rank.

DistSim takes over these sub-models before frameworks actually deploy them, and parses all computation and communication operators. Then DistSim gathers all identical operators to a data structure called \textit{event}, \revise{including computation events and communication events according to the name of these operators. Events use the operator name, parameters and input shape to distinguish from others. In addition, a supplementary attribute is defined for communication events to show whether the communication is intra-node or inter-node. This attribute is defined by recognizing whether the source and destination rank is on the same node.} Finally, the input model is transferred into a set of events at this phase.

\subsection{Profiling Events}


Computation events, which run \revise{on} one device, can be easily profiled by current profilers such as CUPTI~\cite{CUPTI}. However, communication events are more complicated with multiple devices involved.

In our DistSim, we further divide communication events into point-to-point communication (usually used in activation transmission) and all-reduce communication (usually used in activation gathering and weight synchronization). 
In the following paragraphs, we describe how to profile these two communication events.

\begin{figure}[t]
\centerline{\includegraphics[width=0.9\linewidth]{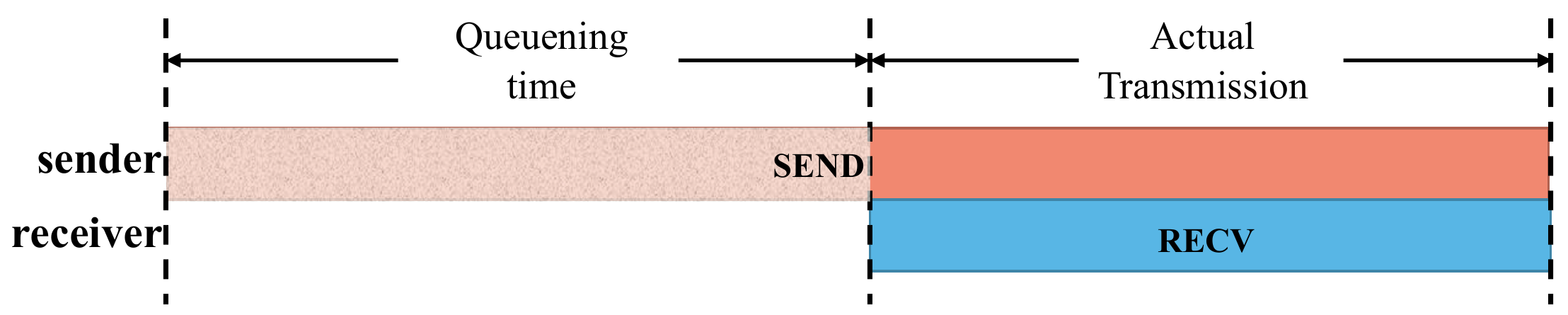}}
\vspace*{-0.3cm}
\caption{Example of the queuing time in the point-to-point communication event. It is not correct to only profile ``SEND'' time.}
\label{queueing_time}
\end{figure}

\vspace*{0.1cm}
\noindent\textbf{Point-to-point Communication Event Profiling.}\hspace*{0.1cm}
The main challenge of profiling point-to-point communication event is that we cannot simply adopt the computation profiling methods into communication because of the queuing time. The transmission cannot be established unless both sender and receiver have called their functions, as shown in Fig.~\ref{queueing_time}.

Based on the observation in Dpro~\cite{dPRO} that the transmission begins when the second function in ``SEND'' and ``RECV'' launches and finishes when both functions exit, the actual transmission time is supposed to be the minimum of ``SEND'' and ``RECV'' calling time. 
Thus, we profile both the sender and receiver during a point-to-point communication event and use the minimum of the two profiling results as the elapsed time of this event.

\vspace*{0.1cm}
\noindent\textbf{All-reduce Communication Event Profiling.}\hspace*{0.1cm}
\revise{Unlike point-to-point communication events, the number of devices participating in all-reduce communication can be greater than 2. When the number of devices is large, it becomes difficult to directly profile the process with limited GPUs.}

Prior work ~\cite{ring-allreduce} has shown that for an all-reduce operator with $N$ devices participating in, there will be two-phase transmission: gather and broadcast. 
In the gather phase, input tensor will be split into $N$ pieces so that each device transfers one piece for $N-1$ times to get one partial sum. 
Then in the broadcast phase, each device broadcasts the partial sum $N-1$ times to all other devices. 
Thus, the total transmission amount per device is $2(N-1)\times P/N$ (where $P$ is the size of tensor), which can be deduced from $N$ and is unrelated to device number $N$ when $N$ is large.

\revise{We adopt this conclusion into our all-reduce communication event profiling. When 8 or fewer devices are involved, we directly run and profile the all-reduce operator; when the number is over 8, we profile the all-reduce process for 8 GPUs, and calculate the time using the above formula that takes into account the actual number of GPUs being used. During our evaluation in Sec.\ref{overall-acc}, we find this calculation's effect on predicting iteration time is limited (<2\%).}


\subsection{Hierarchical Modeling}

\begin{figure*}[tbp]
\centerline{
  \subfigure[Bert-Large]{
     \label{bert-overall}
     \includegraphics[width=0.33\linewidth]{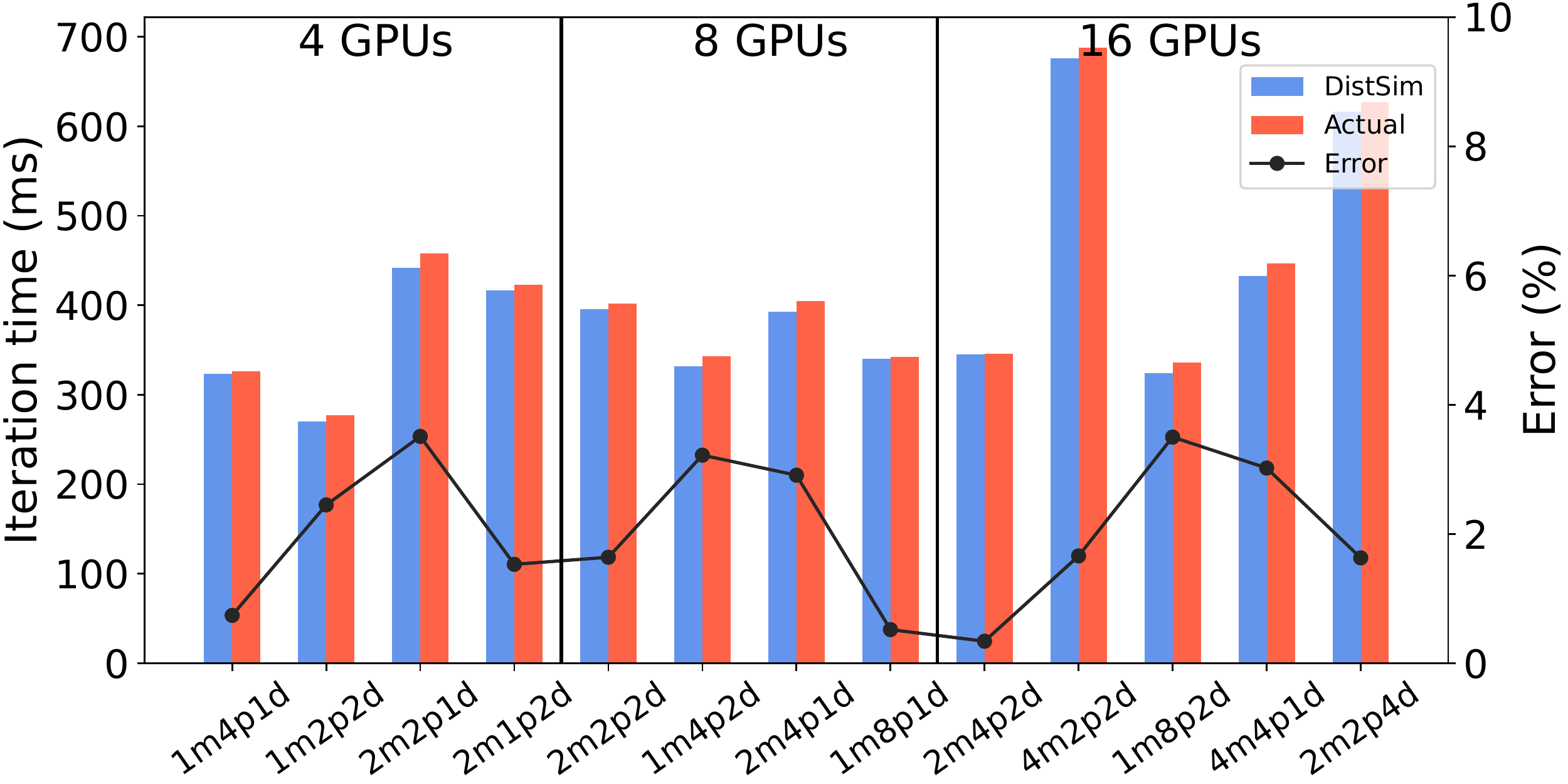}
  }
  \subfigure[GPT-2-345M]{
     \label{GPT-2}
     \includegraphics[width=0.33\linewidth]{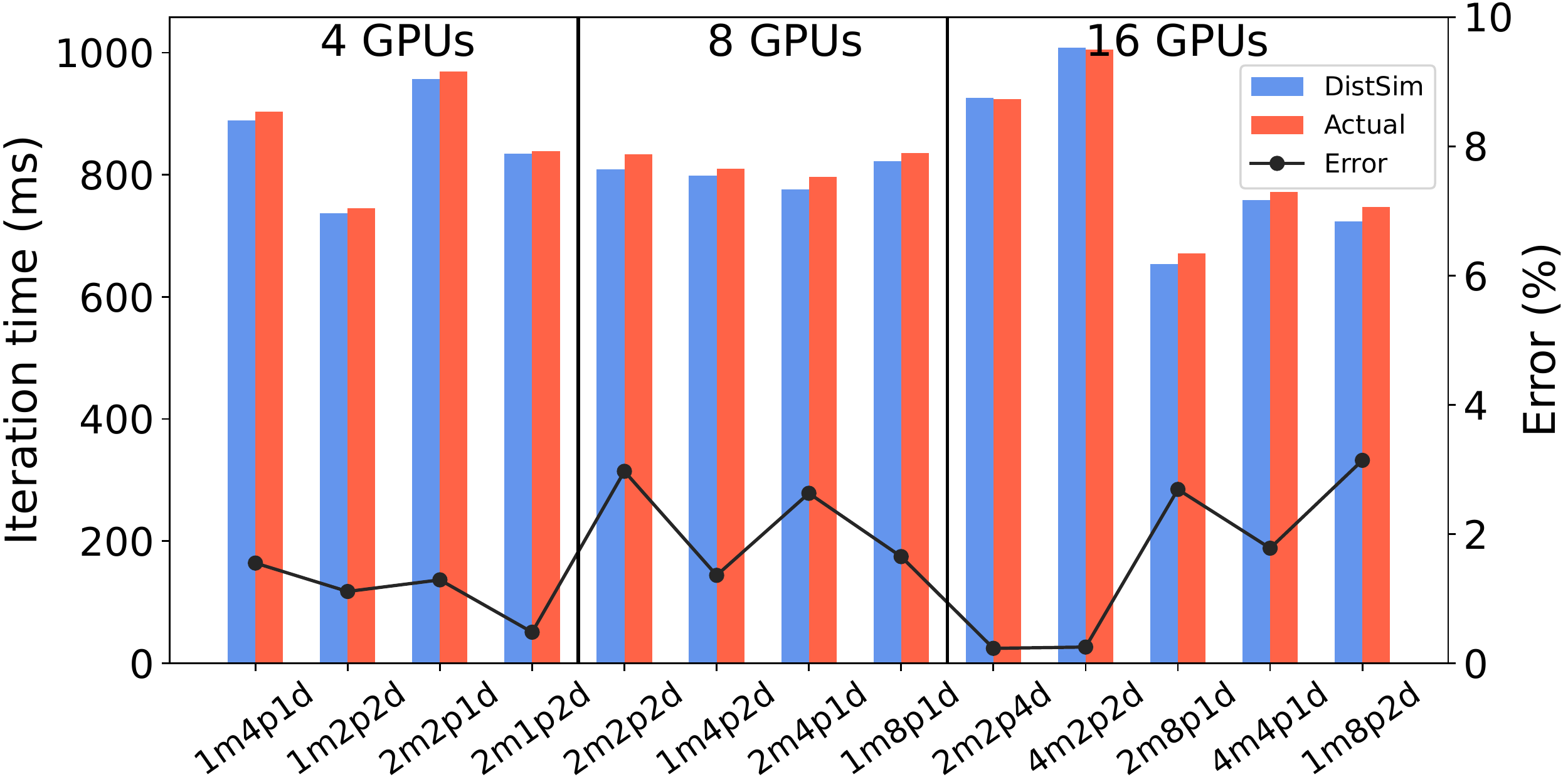}
  }
  \subfigure[T5]{
     \label{T5}
     \includegraphics[width=0.33\linewidth]{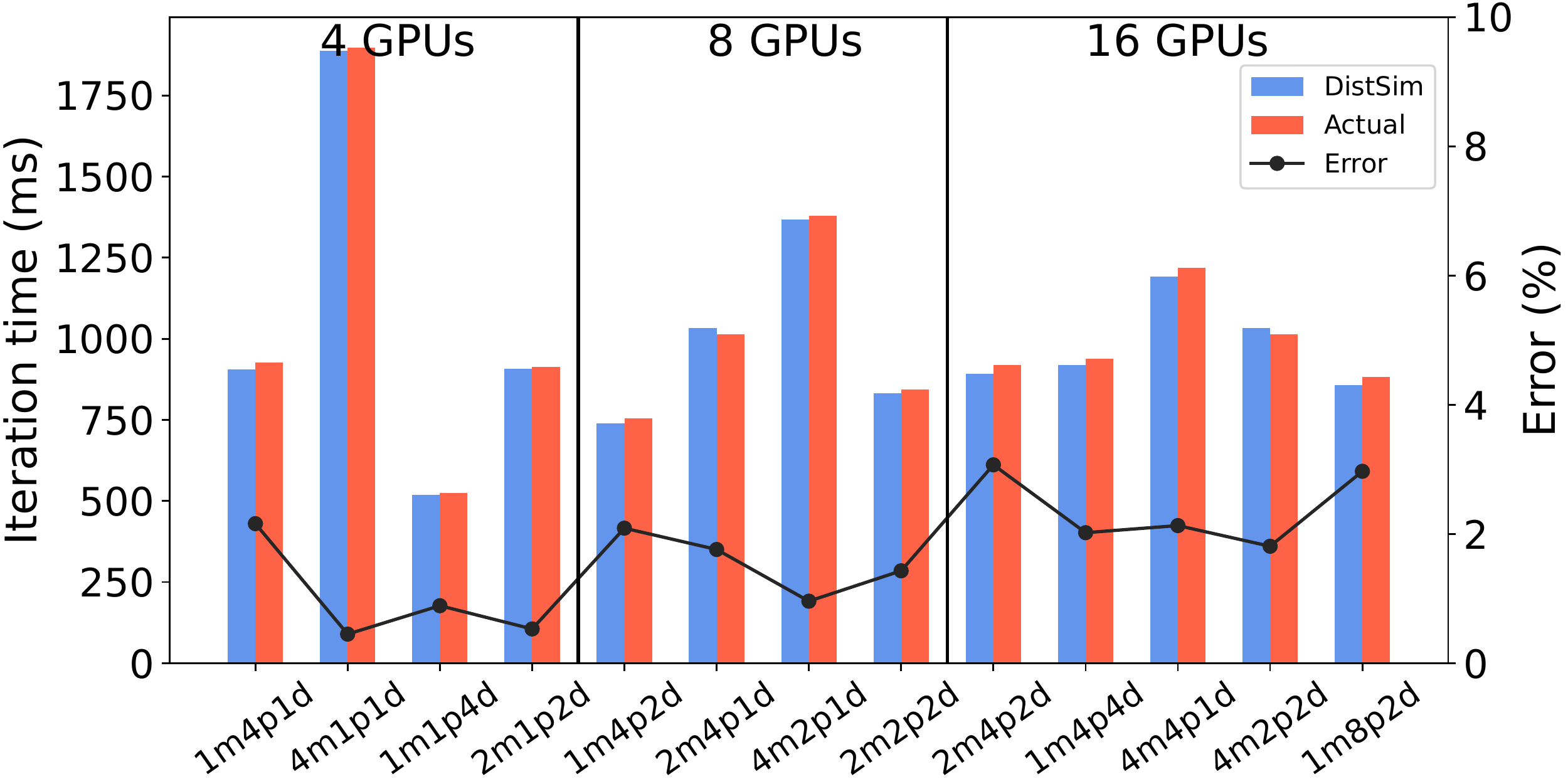}
  }
}
\vspace*{-0.4cm}
\caption{\revise{The evaluation of batch-time (iteration time) between DistSim and actual result on Bert-Large (a), GPT-2-345m (b) and Text-to-Text Transfer Transformer(T5) (c). The x-axis is the strategy of hybrid parallelism distributed training.}}
\vspace*{-0.2cm}
\label{overall-result}
\end{figure*}

\begin{figure*}[tbp]
\vspace*{-0.3cm}
\centerline{
  \subfigure[Bert-Large]{
     \label{bert-detail}
     \includegraphics[width=0.33\linewidth]{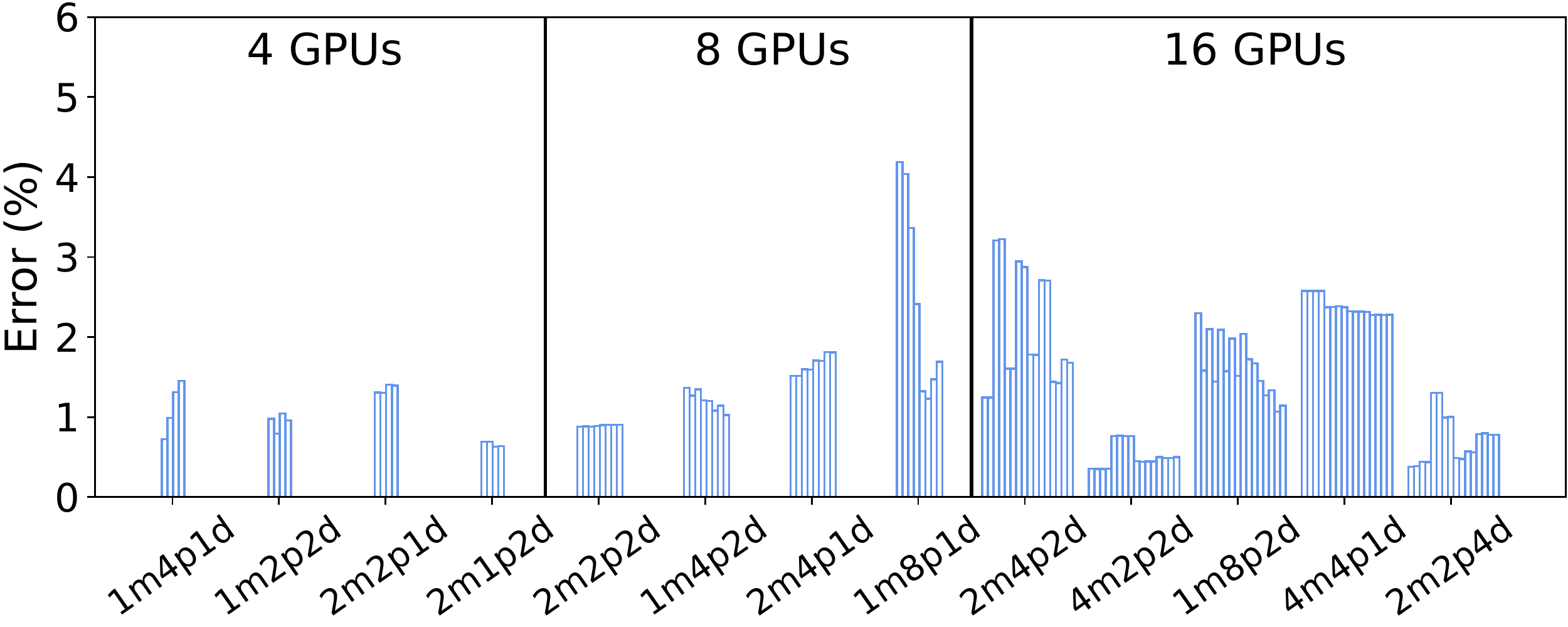}
  }
  \subfigure[GPT-2-345M]{
     \label{GPT-detail}
     \includegraphics[width=0.33\linewidth]{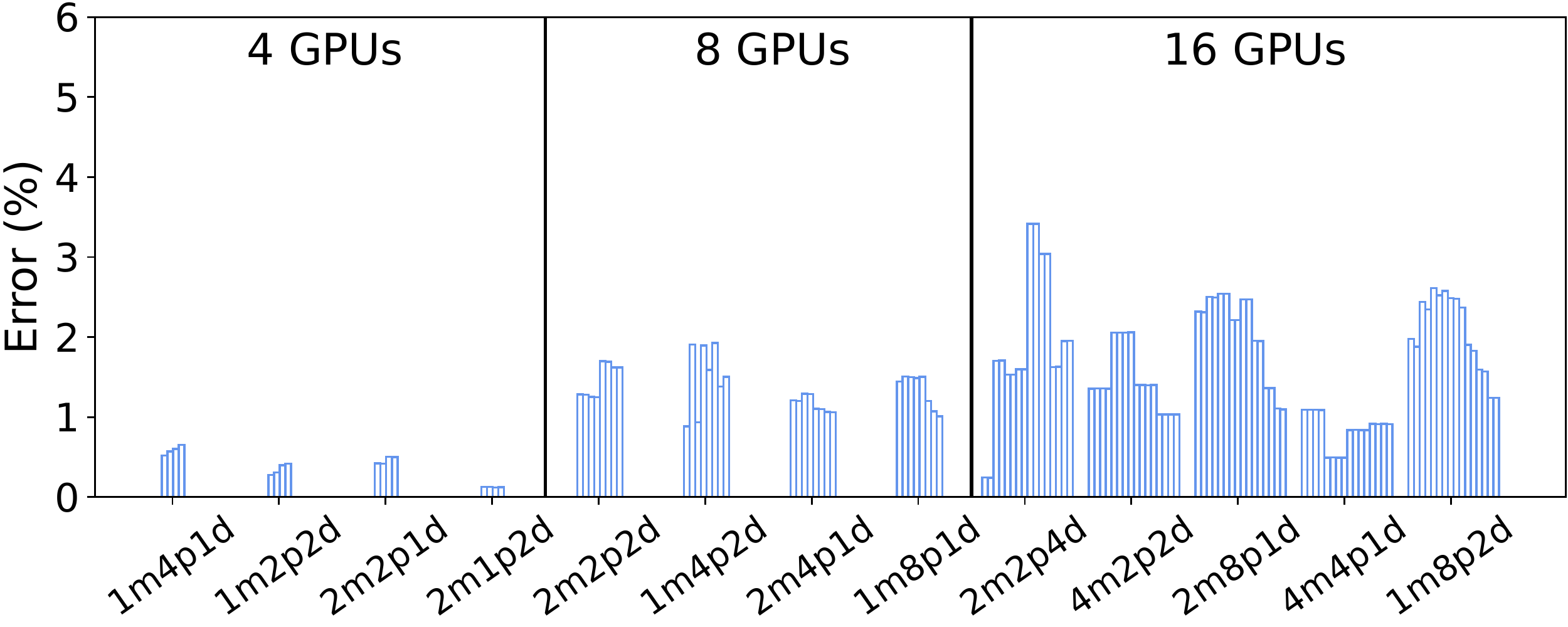}
  }
  \subfigure[T5]{
     \label{T5-detail}
     \includegraphics[width=0.33\linewidth]{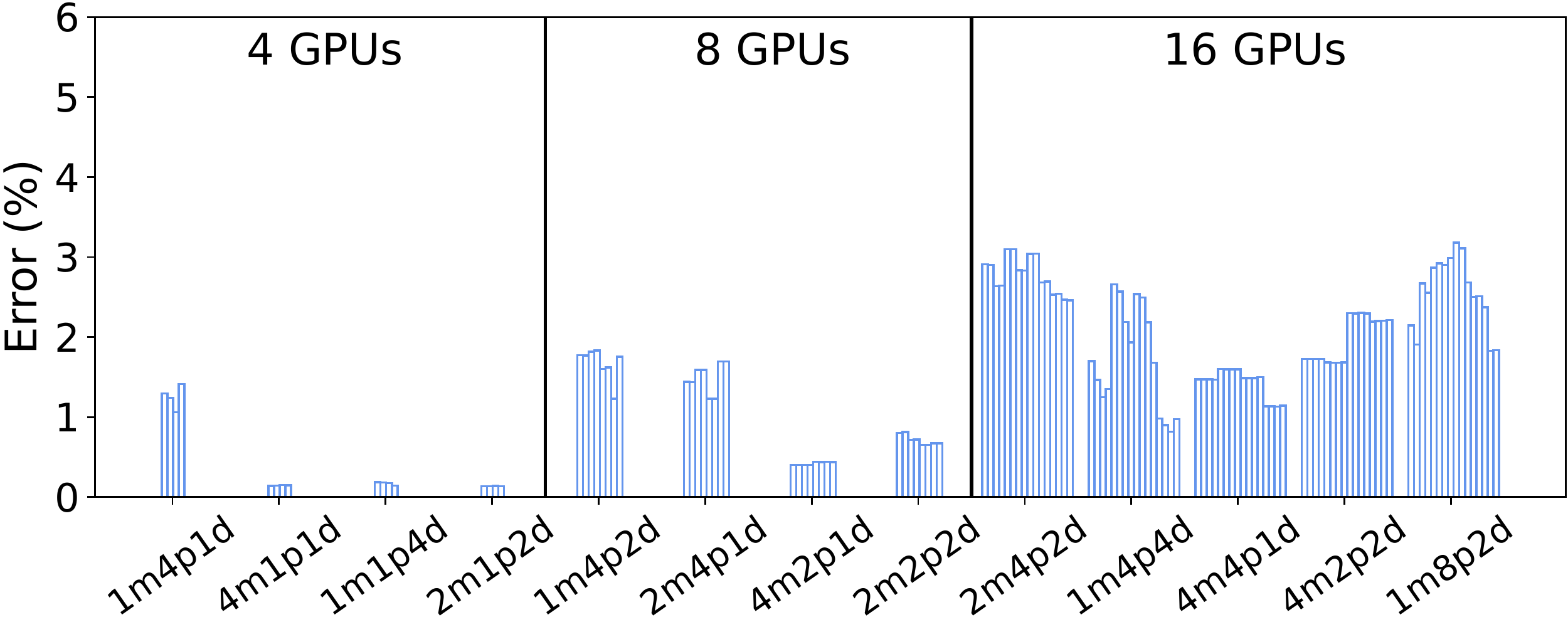}
  }
}
\vspace*{-0.4cm}
\caption{\revise{The evaluation of detailed GPUs' activities between DistSim and actual result on Bert-Large (a), GPT-2-345m (b) and T5(c). The x-axis is the strategy of hybrid parallelism distributed training. Each bar in a certain strategy represents one GPU's activity error.}}
\vspace*{-0.1cm}
\label{detail-result}
\end{figure*}

{
\begin{algorithm}[tbp]
\caption{Pipeline Training Modeling Algorithm}
\label{alg:alg1}
\footnotesize
\LinesNumbered 
\KwIn{Pipeline Scheduling : $S(D,L,M)$, layer-to-events mapping : $MAP(L\rightarrow E)$, model parallelism size : $MP$}
\KwOut{The event-list of pipeline : $F$}
// In $S$, $D/L/M$ represents the set of devices, layers and micro-batches respectively\\
$F \gets LIST[D \times MP]$; // Initialize event-list\\

\While{$S \neq \emptyset$}{
    // Find the first stage in the schedule that matches restrictions, and also return time-stamp to insert the event\\
    $st(d,l,m), t\gets first\_available(S)$\;
    $e\gets MAP[l]$\;
    // Get the point-to-point communication event for the transmission between stages\\
    $e_{comm} \gets get\_comm\_event(st)$\;
    \For{device $m\_p \in [1...MP]$}{
        $F[d\times MP + m\_p] \gets F[d\times MP + m\_p] \cup (e, t, m) \cup (e_{comm}, t + e.elapsed, m)$
    }
    $S\gets S - st$
}
return $F$
\end{algorithm}
}

DistSim would construct the complete training timeline of the training process using model, pipeline, and data parallelism modeling sequentially based on the hierarchical dependencies for different distributed training strategies.

\vspace*{0.1cm}
\noindent\textbf{Model Parallelism Modeling.}\hspace*{0.1cm}
In model parallelism modeling, the set of events and model parallelism size will be used to generate the mapping from layers to events as output. When modeling parallelism is 1, layers will be mapped to a single computation event. Otherwise, the layers will be mapped to a composite event with multiple devices, each containing a computation event and an all-reduce communication event.

\vspace*{0.1cm}
\noindent\textbf{Pipeline Parallelism Modeling.}\hspace*{0.1cm}
The layer-to-events mapping from the model parallelism modeling is used as input, together with the pipeline schedule generated from the pipeline algorithm (currently DistSim has implemented Dapple ~\cite{dapple} and GPipe ~\cite{gpipe}) and micro-batch-size. 
The pipeline scheduling contains the mapping from each events towards the device index and the chronological order of each micro-batch's forward and backward.

The pipeline parallelism modeling follows Algorithm~\ref{alg:alg1} to construct the event-lists of pipeline training. First, the event-flow generator finds the first stage in the schedule which is available (Line~5) (i.e., data is prepared and micro-batch index is matched). The corresponding event is selected and an additional point-to-point communication event is also added if the stage needs to transfer activation to the next stage (Line~6-8). Then these events will be added to all the devices participating in computing this stage because of the model parallelism (Line~9-11). The generator will continue to add events into the event-list until the pipeline schedule is traversed.

If no pipeline strategy is involved, the pipeline schedule will become a sequential flow with only one stage, and the pipeline parallelism size will be set to 1.
After pipeline modeling, the generator gets the event-list of $MP\times PP$ devices, where $MP/PP$ represents model parallelism size and pipeline parallelism size, respectively.

\vspace*{0.1cm}
\noindent\textbf{Data Parallelism Modeling.}\hspace*{0.1cm}
The event-list will be expanded from $MP\times PP$ devices into $MP\times PP \times DP$ devices by duplicating all the events $DP$ times, where $DP$ is the data parallelism size. Additionally, an all-reduce communication event will be added at the end of each event-lists according to the gradient size to be reduced.
\section{Evaluation}

To show the accuracy of simulation towards real performance, we evaluate the following metrics:

\begin{itemize}[leftmargin=*]
    \item The time consumption of one iteration in the training process (batch time). It reflects the accuracy of the critical path of training and \revise{can be used to evaluate the performance of certain strategies or estimate the whole training time given a certain number of training epochs.}(Sec.~\ref{overall-acc})
    \item The timestamps of the beginning and the ending of computation events. It represents the consistency of each device's activity, \revise{which helps users to apply more computations when one GPU is idle to achieve more utilization.} (Sec.~\ref{detailed-acc})
    \item A detailed comparison of per-stage timestamps. In this metric, we show the differences for \textit{every pipeline stage} and \textit{every micro-batch} between DistSim and actual running, which shows similarities between the modeled timeline and actual timeline. \revise{It helps programmers to locate pipeline bubbles and performs practical operations such as fault-tolerance~\cite{qiu2019adversarial, checkfreq, resilience} during bubbles.}(Sec.~\ref{perstage-acc})
    \item We also evaluate DistSim to model the training for a large-scale network with 128 GPUs, compared with Megatron-LM\cite{megatron-exp}. (Sec.~\ref{generalization})
\end{itemize}

\subsection{Experiment Setup \label{exp-setup}}

\vspace*{0.1cm}
\noindent\textbf{Testbed.}\hspace*{0.1cm} We evaluate the accuracy of DistSim in a cluster with up to 16 Nvidia-A40 GPUs on 4 servers. We use the whole 16 GPUs for real distributed training performance tracing, and 4 GPUs with 2 servers for DistSim profiling and simulation.
We use PyTorch-Distributed with CUDA 11.6 and NCCL backend as our evaluation framework. We leverage the ability of model partition and generation in Megatron-LM ~\cite{megatron} as our model partitioner in DistSim.

\vspace*{0.1cm}
\noindent\textbf{Benchmarks.}\hspace*{0.1cm}
We train three models, BERT-Large ~\cite{bert}, GPT-2-345M ~\cite{gpt2} and Text-to-Text Transfer Transformer(T5) ~\cite{T5}, on 1, 2 and 4 servers with 4, 8, 16 GPUs as our evaluation workload. We select various distributed training strategy configurations to evaluation, denoted as ``xM xP xD'', which means the size of each dimension of parallelism (model, pipeline, and data parallelism accordingly).

\subsection{Overall Accuracy Evaluation \label{overall-acc}}

We evaluate DistSim's prediction accuracy of iteration time, called overall batch-time evaluation. We first generate event-flow with DistSim and calculate the analyzed iteration time. Then we profile on a real cluster to get the accurate iteration time. We compare the simulated and actual batch-time and analyze the error.

We apply the evaluation to various hybrid parallelism strategies and three different models described in the benchmarks. The evaluation result shows in Fig.~\ref{overall-result}. We find that DistSim evaluates the iteration time in high accuracy, with at most $3.51\%$ errors, much lower than analytical approach presented in Fig.~\ref{res_heuristic}.

During the evaluation, we notice that the error has weak relationship with the number of GPUs and the parallelism strategies. For example, in Bert-Large training, the error in less GPUs (``1M2P2D'', 4 GPUs, 2.45\%) is larger than more GPUs (``2M2P4D'', 16 GPUs, 1.63\%). We address this phenomenon as the random fluctuation during profiling since we only profile 100 iterations to get the average batch-time.

\begin{figure*}[tbp]
\centerline{\includegraphics[width=0.8\linewidth]{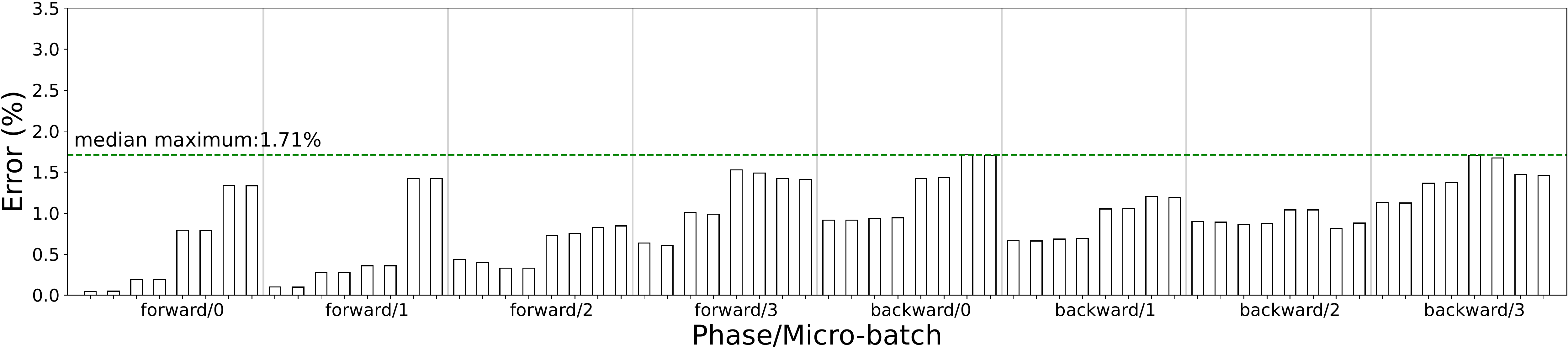}}
\vspace*{-0.3cm}
\caption{\revise{The detailed per-stage comparison result for Bert\cite{bert} model with model parallelism size 2 and pipeline parallelism 4. The micro-batch size is set to 4. The x-axis is the training phase (forward, backward) of a certain micro-batch and the y-axis is the median error for the DistSim simulation and 100-times actual running. Each bar in one stage represents the result of GPU0-7, respectively.}}
\label{res-per-event}
\end{figure*}

\subsection{Per-GPU Detailed Accuracy Evaluation \label{detailed-acc}}

We evaluate DistSim's simulation result of every GPU. In the event-flow generated by DistSim, we select all the events' beginning and ending timestamps and calculate the average bias from the actual timeline. The results in Fig.~\ref{detail-result} show high accuracies of the GPU-level modeling, with at most $4.19\%$ errors.

We find that some GPUs' errors are higher than the average level (e.g., GPU 9-12 in GPT-2 ``2M2P4D'' training). There are mainly three reasons. First, we find during our evaluation that the profiling fluctuation affects individual devices' activities more than batch-time since the fluctuation will accumulate and affect other GPUs. Second, we use rank 0's clock time as a global standard for simplification. Therefore, it introduces the time alignment problem addressed in dPRO~\cite{dPRO}, and the actual timestamp should be adjusted. Third, there is no device synchronization in data parallelism until weight all-reduce, so there \revise{exist} timeline variations among different data parallelism ranks. 

\revise{Additionally, our evaluation has shown that when more GPUs are involved, the margin of error seems to increase. After further analysis, we find that the error positively correlates with the pipeline parallelism size. This is because errors in earlier pipeline stages can impact later stages, resulting in a greater likelihood of deviations from the analyzed model's timeline in larger pipeline parallelism scenarios.}

\subsection{Per-stage Accuracy Evaluation \label{perstage-acc}}

We also evaluate after DistSim's modeling, how similar for each forward and backward computation is between DistSim and the golden result. We use the parallel configuration ``2m4p1d'' and micro-batch size 4, therefore it will include 32 forward and backward stages in total, with 4 per GPU. The difference between the start and finish timestamps according to the whole training is regarded as the error. We run the real training iteration 100 times, and the error distribution is presented in Fig.~\ref{res-per-event}.

In the result, we find that the largest medium error among every stage and every GPU is 1.71\%, which shows the low bias from the simulation and actual running. Additionally, the error distribution for every two GPUs (0 and 1, 2 and 3, etc.) is generally the same, which shows the computation of different model parallelism parts can be regarded as the same. We also find interestingly that only the error in the first stage of the first 2-4 GPUs is approaching 0 and others are mainly equal. This is because we use the start timestamp of the first stage as the global standard time for both simulated and actual timelines. Therefore, the error for this timestamp is 0.

\subsection{Large-scale Generalization Evaluation \label{generalization}}

\begin{figure}[tbp]
\centerline{\includegraphics[width=0.7\linewidth]{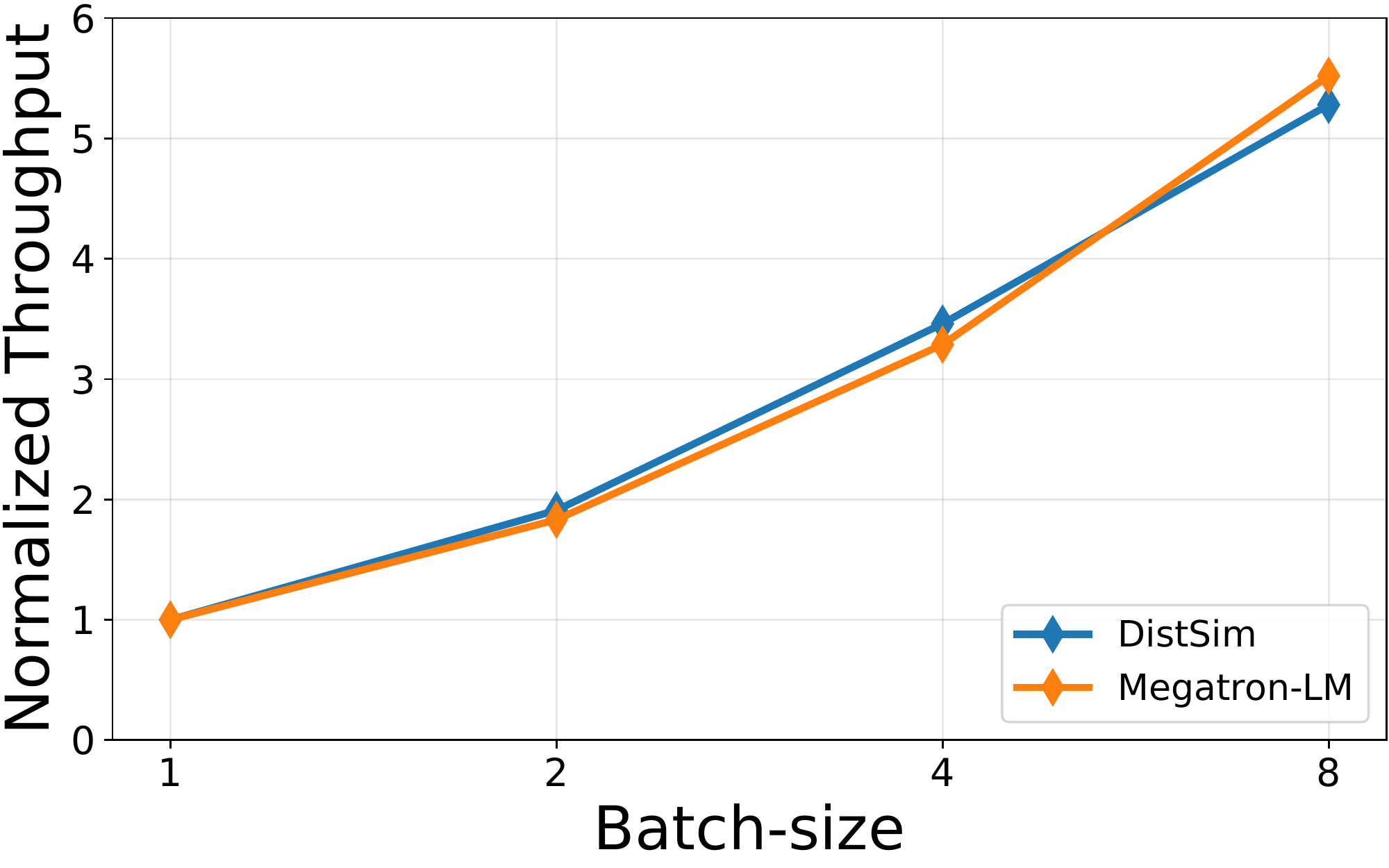}}
\vspace*{-0.2cm}
\caption{A thoughput comparison between modeling by DistSim and actual running reported by Megatron-LM on 128 GPUs, training for a GPT model with 145 billion parameters using model parallelism size 8 and pipeline parallelism 16. The reported data can be found in Fig.17 in the experiments of Megatron-LM~\cite{megatron-exp}.}
\vspace*{-0.2cm}
\label{res-large}
\end{figure}

To evaluate how DistSim performs on a real large-scale model, we use DistSim to model the training process of a 145-billion-parameter GPT model with 128 GPUs. The distributed configuration is ``8M16P1D'', which is given by Megatron-LM ~\cite{megatron-exp}.

We compare the modeling result with data reported by Megatron-LM ~\cite{megatron-exp}. Because the profiling hardware and the bandwidth is different with Megatron-LM, we only compare the normalized throughput according to batch-size 1 to evaluate whether the modeling is in accordance with fact. The result is in Fig.~\ref{res-large}. We find that the throughput increment rate has high similarities between DistSim and Megatron-LM, which shows the ability of DistSim to model large-scale training.
\section{Use-case: Auto Parallel Strategy Search}

In this section, we provide a simple yet important use-case of DistSim. Since DistSim can accurately evaluate the training process's throughput with different hybrid parallelism strategies, it can be used to find an optimal strategy before using the actual cluster.

The strategy search is applied on new unseen model called ``BERT-exLarge`` with 48 transformer layers on 4 new nodes with 16 A10 GPUs. Suppose the global batch-size is fixed into 16, then the throughput evaluation can be regarded as iteration time.

We use a grid-search method to traverse the hybrid parallelism search space with the help of DistSim. There are 5 configuration choices for each of the parallelism dimension -- 1, 2, 4, 8 and 16. Overall, there are 15 different hybrid parallelism settings as some combinations are invalid. We only need to traverse along the pipeline- and the model- parallelism axis as data parallelism size can be conducted from the above two ($DP = GPUs / MP / PP$).

\begin{figure}[tbp]
\centerline{\includegraphics[width=0.75\linewidth]{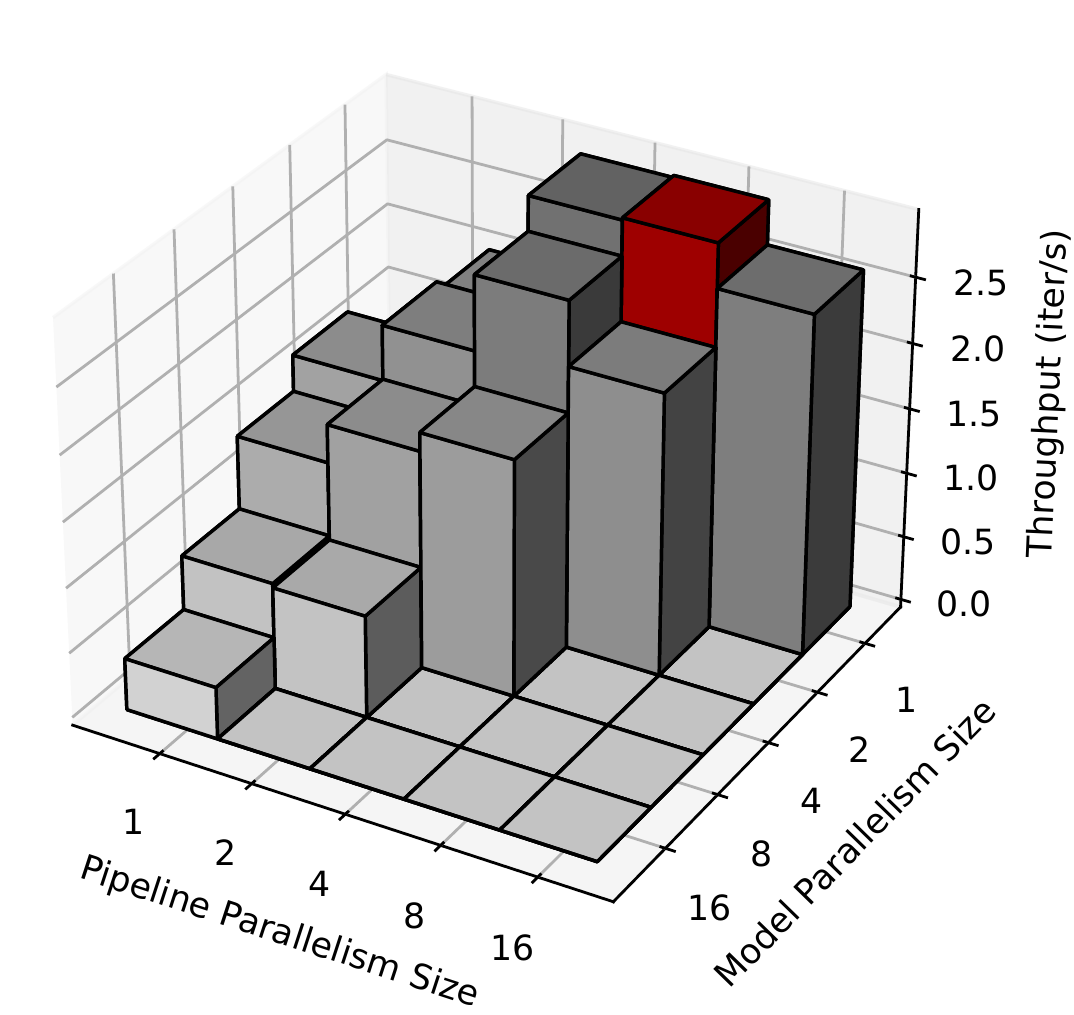}}
\vspace*{-0.1cm}
\caption{BERT-exLarge training strategy grid-search results  using DistSim on 4 nodes with 16 GPUs. The red bar (data and pipeline parallelism size of 2 and 8) achieves the best throughput. Note that the unreachable configurations are drawn to 0.}
\vspace*{-0.2cm}
\label{grid-search}
\end{figure}

We evaluate all the settings with DistSim, and show the results  in Fig.~\ref{grid-search}. The result indicates that the strategy whose data parallelism size 2 and pipeline parallelism size 8 is optimal with throughput over 2.94 iterations per second, which at most increases 7.37$\times$ from the worst strategy (model parallelism size 16).

Next, we run on an actual 16 GPUs cluster to verify whether the searched strategy is reliable. Table.~\ref{tab:real-analyze} shows high similarities between DistSim's grid search and the accurate profiling result.

\revise{
In addition, we also analyze the cost of time in profiling and simulation compared to direct running, which shows in Table.~\ref{tab:time_cost}. We find that DistSim only takes about 12.96\% of time compared to direct running due to the reduction of profiling redundancy. The main cost in profiling is some all-reduce operators since their redundancy is limited. For example, in ``2m1p8d'' strategy, there are only two same all-reduce operators with 8 GPUs involved, and we can only reduce 50\% profiling redundancy. More redundancy could be discovered and reduced on a larger scale. Also, from the result, we find that simulation time only occupies a tiny portion of the total time cost (<1\%), so if the profiling time can be reduced, the total cost could be further eliminated.
}


\section{Related Work And Discussion}

\vspace*{0.1cm}
\noindent\textbf{Distributed Training Performance Modeling.}\hspace*{0.1cm}
DistIR\cite{distIR} defines a distributed intermediate representation (IR) to represent the computation and communication flow of distributed training, aiming to reveal the behavior of each rank (device) and deploy each rank's work more conveniently. However, DistIR is simply an analytical model, which shows unsatisfying accuracy \revise{(over 30\% on average)} between actual training and simulated training, limiting its usage for optimizing the performance and throughput. 
Daydream~\cite{Daydream} and dPRO~\cite{dPRO} both focus on using a profiler and simulator to evaluate the performance and bottleneck in distributed data-parallel training. Daydream~\cite{Daydream} proposes a kernel-level profiler based on CUPTI~\cite{CUPTI} to get a kernel-level time and maps kernel to model operators to get the operator computation and communication cost. Then they use a simulator to replay the training process and identify how much throughput improves when using a certain optimization method. 
Similarly, dPRO~\cite{dPRO} leverages the frameworks' profiling ability (TensorFlow or PyTorch) to get the operator-level time. They also propose a global data-flow graph (DFG) to simulate and replay the training process when certain optimization of distributed training is used. However, those approaches only considered data parallelism in distributed training. Their methods fail to analyze the situation when a model is too large to deploy on a single device and other distributed training strategies (e.g., pipeline or model parallelism) must be imported.

\begin{table}[t]
\Huge
\centering
\vspace*{-0.2cm}
\caption{\revise{Comparison between DistSim-based distributed training strategy grid-search and actual measurement.}}
\vspace*{-0.3cm}
\label{tab:real-analyze}
\resizebox{\columnwidth}{!}{%
\begin{tabular}{|c|c|c|c|c|}
\hline
        & \begin{tabular}[c]{@{}c@{}}best strategy\\ (iter/s)\end{tabular} & \begin{tabular}[c]{@{}c@{}}second-best \\ strategy\\ (iter/s)\end{tabular} & \begin{tabular}[c]{@{}c@{}}worst strategy\\ (iter/s)\end{tabular} & speed up \\ \hline
DistSim & 2.94                                                             & 2.92                                                                    & 0.398                                                             & 7.379$\times$    \\ \hline
Actual  & 2.97                                                             & 2.90                                                                    & 0.396                                                             & 7.488$\times$    \\ \hline
\end{tabular}%
}
\end{table}

\begin{table}[]
\Huge
\centering
\caption{\revise{The comparsion of time cost between DistSim and actual run while searching for the best strategy.}}
\vspace*{-0.3cm}
\label{tab:time_cost}
\resizebox{\columnwidth}{!}{%
\begin{tabular}{|c|c|c|c|}
\hline
    & \revise{Simulate Time (s)} & \begin{tabular}[c]{@{}c@{}}\revise{Profiling GPU}\\ \revise{Time (gpu$\times$s)}\end{tabular} & \revise{Relative Scale} \\ \hline
\revise{DistSim}                                              & \revise{0.14}                                                       & \revise{49.18 }                                                             & \revise{0.1296$\times$}                                                   \\ \hline
\revise{Direct Run} & \revise{- }                                                          & \revise{380.35}                                                             & \revise{1$\times$}                                                      \\ \hline
\end{tabular}%
}
\end{table}

\vspace*{0.1cm}
\noindent\textbf{Modeling in Auto Parallel Strategy Search.}\hspace*{0.1cm}
To find the optimal parallel strategy, many works, such as Piper~\cite{piper}, AccPar~\cite{accpar}, PaSE~\cite{pase} and Alpa~\cite{alpa}, construct their cost models toward the training throughput and propose algorithms heuristically to search the optimal. Piper, PaSE and Alpa define cost models according to the latency of training iteration. Then they use dynamic programming approaches to search for the least latency according to their different search spaces. AccPar~\cite{accpar} uses a layer-wise recursive algorithm to find the optimal partition of each layer in the data and model parallelism domain. Although these approaches can find a state-of-the-art distributed training strategy, \revise{they are} unable to evaluate how much throughput improves and why their approaches outperform from devices' activity aspect unless evaluated on real clusters. Moreover, since most of the works use analytical assumptions~\cite{piper,accpar,pase}, such as ``computing time is equal to the division of operator count and computing capacity'', the result of the cost models only shows the advantage of one strategy towards other strategies, but cannot report their detailed statistics.

\revise{
\vspace*{0.1cm}
\noindent\textbf{Discussion: Scalability towards New Strategies, Algorithms, Communication Operators, Models, etc.}\hspace*{0.1cm}
It is important to analyze whether DistSim can handle unseen distributed training configurations. For new distributed strategies such as ZeRO-DP~\cite{Zero} and 3D-parallelism~\cite{3D-parallel}, DistSim will take these strategies as input. Their dependencies can be recognized. For example, ZeRO-DP has a combination of intra-layer and inter-model dependencies. Therefore, the model can be partitioned from these strategies, DistSim can generate events and perform modeling. For new algorithms such as asynchronous pipeline parallelism like Pipedream~\cite{pipedream}, the schedule in pipeline parallelism modeling can still be established only without a global synchronize event (usually an all-reduce event), so pipeline modeling can still complete. For new communication or computation operators in unseen models, DistSim can regard them as new events and perform profiling normally.
}

\section{Conclusion}

It is important to evaluate and analyze the performance of throughput and utilization in distributed DNN training. DistSim is a new performance model that can achieve high accuracy with low profiling cost in combining data, model and pipeline parallelism. DistSim achieves this by two key \revise{contributions}: (1) the same computation and communication processed by different devices, resulting in profiling redundancy; (2) the hierarchy of different parallelism strategies with different partition granularity, which makes it possible to model the training process step by step. Our evaluation shows that DistSim achieves high accuracy in modeling the training process with different strategy configurations and can be used in various tasks such as training strategy autotuning.

\begin{acks}
This work was supported by the National Key R\&D Program of China under Grant 2022YFB4501401, the National Natural Science Foundation of China (NSFC) grant (62222210, and 62072297, and 61832006).
The authors would like to thank the anonymous reviewers for their constructive feedback for improving the work. 
Any opinions, findings, and conclusions in this paper are those of the authors only and do not necessarily reflect the views of our sponsors.
\end{acks}

\bibliographystyle{ACM-Reference-Format}
\balance
\bibliography{acmart}

\end{document}